\documentclass[aps,prb,twocolumn,showpacs,preprintnumbers,amsmath,amssymb,superscriptaddress]{revtex4}%

\usepackage{graphicx}%
\usepackage{dcolumn}
\usepackage{amsmath}
\usepackage{color}

\makeatletter
\def\btt#1{\texttt{\@backslashchar#1}}%
\DeclareRobustCommand\bblash{\btt{\@backslashchar}}%
\makeatother

\begin{document}


\title{Zero-field superfluid density in $d-$wave superconductor evaluated from the results
of muon-spin-rotation experiments in the mixed state}

\author{R.~Khasanov}
 \email{rustem.khasanov@psi.ch}
 \affiliation{Laboratory for Muon Spin Spectroscopy, Paul Scherrer
Institut, CH-5232 Villigen PSI, Switzerland}
\author{Takeshi~Kondo}
 \affiliation{Ames Laboratory and Department of Physics and Astronomy, Iowa State University, Ames, IA~50011, USA}
 \affiliation{Department of Crystalline Materials Science, Nagoya University, Nagoya 464-8603, Japan}
\author{S.~Str\"assle}
 \affiliation{Physik-Institut der Universit\"{a}t Z\"{u}rich,
Winterthurerstrasse 190, CH-8057 Z\"urich, Switzerland}
\author{D.O.G.~Heron}
 \affiliation{School of Physics and Astronomy, University of St. Andrews, Fife,
KY16 9SS, UK}
\author{A.~Kaminski}
 \affiliation{Ames Laboratory and Department of Physics and Astronomy, Iowa
State University, Ames, IA~50011, USA}
\author{H.~Keller}
 \affiliation{Physik-Institut der Universit\"{a}t Z\"{u}rich,
Winterthurerstrasse 190, CH-8057 Z\"urich, Switzerland}
\author{S.L.~Lee}
 \affiliation{School of Physics and Astronomy, University of St. Andrews, Fife,
KY16 9SS, UK}
\author{Tsunehiro Takeuchi}
\affiliation{Department of Crystalline Materials Science, Nagoya
University, Nagoya 464-8603, Japan} \affiliation{EcoTopia Science
Institute, Nagoya University, Nagoya 464-8603, Japan}

\begin{abstract}
We report on  measurements of the in-plane magnetic penetration $\lambda_{ab}$ in the optimally doped cuprate superconductor (BiPb)$_2$(SrLa)$_2$CuO$_{6+\delta}$ (OP Bi2201) by means of muon-spin rotation ($\mu$SR). We show that in unconventional $d-$wave  superconductors (like OP Bi2201), $\mu$SR experiments conducted in various magnetic fields allow to evaluate the zero-field magnetic penetration depth $\lambda_0$, which relates to the zero-field superfluid density in terms of $\rho_s\propto\lambda_0^{-2}$.
\end{abstract}

\maketitle


Muon-spin-rotation ($\mu$SR) measurements in  the mixed state of
type-II superconductors provide valuable information on the
superconducting properties. An important
advantage of this method is that the muons probe the {\it bulk} of the
material, and the results are not complicated by surface
imperfections. The quantitative parameters extracted from $\mu$SR
experiments depend, however, on the details of the model applied to
reconstruct the internal magnetic field distribution in the superconductor in the mixed state. So far, field distributions
measured by means of $\mu$SR were analyzed within the framework of analytical
models based on London and Ginzburg-Landau (GL) theories, which can be
applied, in general, to conventional superconductors with a
single isotropic energy gap.\cite{Brandt88,Yaouanc03,Brandt03} The
situation becomes much more complicated in the case of
unconventional superconductors, like cuprates, MgB$_2$, {\it etc.} It was found, in particular,
that the effective magnetic field penetration depth $\lambda_{eff}$ extracted from
$\mu$SR measurements depends on the applied magnetic field ( see
{\it e.g.}
Refs.~\onlinecite{Sonier99,Sonier00,Kadono04,Niedermayer02,Serventi04,
Khasanov07_La214,Khasanov07_Y124,Khasanov07_Y123}) which is
unexpected within the GL theory. Here $\lambda_{eff}$ refers to a quantity evaluated from $\mu$SR experiments conducted in a superconductor in the mixed state, in contrast to $\lambda_0$ as obtained from Meissner state experiments ($H\ll H_{c1}$, $H_{c1}$ denotes the lower critical field). In addition, it was observed that not
only the absolute value, but also the shape of $\lambda^{-2}_{eff}(T)$
changes with field.
\cite{Sonier99,Khasanov07_La214,Khasanov07_Y124,Khasanov07_Y123,Niedermayer02}
In this respect the question concerning the relation of $\lambda_{eff}$ to $\lambda_0$, which is generally assumed to be proportional to the superfluid density ($\lambda_0^{-2}\propto\rho_s$), becomes very important.

In this paper we report on the results of a $\mu$SR
study of the in-plane magnetic penetration depth $\lambda_{ab}$ in
optimally doped (BiPb)$_2$(SrLa)$_2$CuO$_{6+\delta}$.
$\lambda_{eff}(T,H)$ was obtained from the measured temperature
dependence of the $\mu$SR linewidth by using numerical calculations
of Brandt.\cite{Brandt03} The temperature dependence of $\lambda_0^{-2}$ was further evaluated from $\lambda_{eff}(T,H)$
considering the nonlinear and the nonlocal response of a
superconductor with  nodes in the energy gap to the applied magnetic
field. It was found that only at relatively low magnetic fields
[$B/B_{c2}(0)\lesssim 10^{-3}$, $B_{c2}(0)$ is the zero-temperature
value of the upper critical field] $\lambda_{eff}$ is a good measure of $\lambda_0$. The high field data,
however, need to be evaluated by taking into account both the
nonlinear and the nonlocal corrections.

\begin{figure}[htb]
\includegraphics[width=0.9\linewidth]{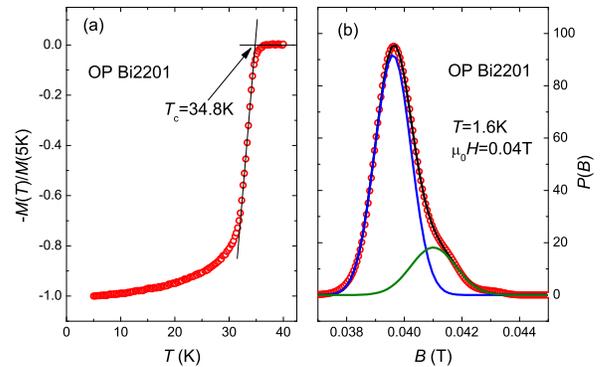}
 \vspace{-0.3cm}
\caption{(Color online) (a) Field-cooled magnetization $M_{FC}(T)$ of
OP Bi2201. The field $\mu_0H=1$~mT was applied parallel to the crystallographic $c$ axis. (b) The magnetic field distribution $P(B)$ of OP Bi2201
taken at $T=1.6$~K, $\mu_0H=0.04$~T. The lines represent the best
fit within a two-Gaussian approach.}
 \label{fig:magnetization-PB}
\end{figure}

Details on the sample preparation of optimally doped
(BiPb)$_2$(SrLa)$_2$CuO$_{6+\delta}$  (OP Bi2201)
single crystals can be found elsewhere.\cite{Kondo04,Kondo05}
Field--cooled magnetization ($M_{FC}$) measurements of OP Bi2201
were performed with a SQUID magnetometer at $\mu_0H=1$~mT,
applied parallel to the $c$ axis, for
temperatures ranging from  $5$~K to $50$~K. The transition
temperature $T_c=34.8$~K was obtained as the intersect of the
linearly extrapolated $M_{FC}(T)$ curve in the vicinity of
$T_c$ with the $M=0$ line [see Fig.~\ref{fig:magnetization-PB}~(a)].

The transverse-field $\mu$SR experiments were carried out at the
$\pi$M3 beam line at the Paul Scherrer Institute (Villigen,
Switzerland).
Two OP Bi2201 single crystals  with an approximate  size of
4$\times$2$\times$0.1~mm$^3$ were mounted on a holder specially
designed to perform $\mu$SR experiments on thin single crystalline
samples. The sample was field cooled from above $T_c$ to 1.6~K in
series of fields ranging from 5~mT to 0.64~T. The magnetic field
was applied parallel to the $c$ axis and
transverse to the muon-spin polarization. The typical counting
statistics were $\sim15-18$ million muon detections per data
point. The experimental data were analyzed within the
same scheme as described in
Refs.~\onlinecite{Khasanov07_La214, Khasanov07_Y124, Khasanov08_Bi2201}.
This is based on a two-component Gaussian fit of the $\mu$SR time
spectra which allows to describe the asymmetric local magnetic field
distribution $P(B)$ in the superconductor in the mixed state [see Fig.~\ref{fig:magnetization-PB}~(b)]. The magnetic
field penetration depth $\lambda$ was derived from the second moment
of $P(B)$ as $\sigma^2\propto\lambda^{-4}$.\cite{Brandt88} The superconducting
part of the square root of the second moment
($\sigma_{sc}\propto\lambda^{-2}$ ) was obtained by subtracting the
normal state nuclear moment contribution ($\sigma_{nm}$) from the
measured $\sigma$, as $\sigma_{sc}^2=\sigma^2-\sigma_{nm }^2$ (see
Ref.~\onlinecite{Khasanov07_La214} for details).
Since the
magnetic field was applied along the crystallographic $c$ axis,
our experiments provide direct information on $\lambda_{ab}$.

\begin{figure}[htb]
\includegraphics[width=0.8\linewidth]{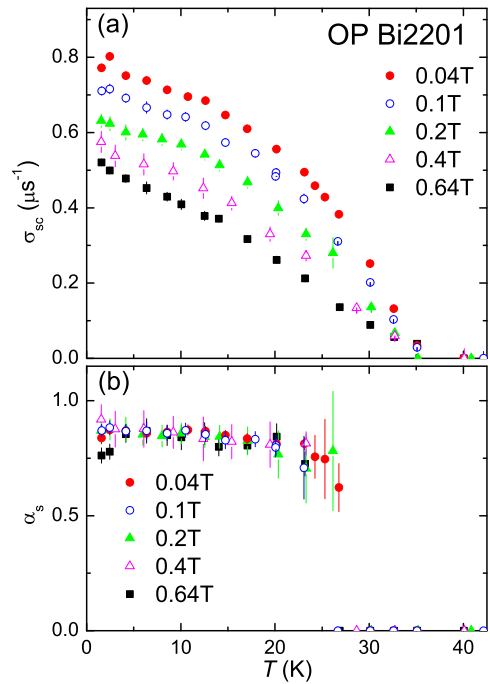}
\caption{(Color online) (a) Temperature dependence of
$\sigma_{sc}\propto\lambda_{ab}^{-2}$ of
OP Bi2201 measured at $\mu_0H=0.04$~T, 0.1~T,
0.2~T, 0.4~T, and 0.64~T. (b) Dependence of the skewness
parameter $\alpha_{s}$ on temperature. }
 \label{fig:sigma_vs_T}
\end{figure}

The  temperature  dependences of
$\sigma_{sc}\propto\lambda_{ab}^{-2}$ measured after field-cooling
the sample from far above $T_c$ in $\mu_0H$=0.04~T, 0.1~T, 0.2~T,
0.4~T, and 0.64~T are shown in Fig.~\ref{fig:sigma_vs_T}~(a). To
ensure that $\sigma_{sc}(T)$ is determined primarily by the variance
of the magnetic field due to the vortex lattice (VL) we plot in
Fig.~\ref{fig:sigma_vs_T}~(b) the corresponding values of the
skewness parameter $\alpha_{s}=\langle \Delta
B^{3}\rangle^{1/3}/\langle \Delta B^{2}\rangle^{1/2}$ [$\langle
\Delta B^{n}\rangle$ is the $n-$th central moment of $P(B)$]. $\alpha_{s}$ is a dimensionless measure of the
asymmetry of the lineshape, the variation of which reflects
underlying changes in the vortex structure.\cite{Lee93}  For an ideal triangular
VL $\alpha_{s}\simeq1.2$. It is very sensitive to structural
changes of the VL which can occur as a function of
temperature and/or magnetic field.\cite{Lee93,Aegerter98}
Fig.~\ref{fig:sigma_vs_T}~(b) implies that in OP Bi2201
$\alpha_{s}(T,H)$ is almost constant for $1.6$~K$\leq T\leq26$~K and
smaller than the ideal value of 1.2, which is probably
caused by distortions of the VL due to pinning effects. The sharp
change of $\alpha_{s}$ at $T\simeq 30$~K is similar to what was
observed in Bi2212, where it was attributed to VL melting.\cite{Lee93,Aegerter98}
Therefore, we conclude that for temperatures
$1.6$~K$<T\lesssim30$~K the $T$ variation of $\sigma_{sc}$ in
OP Bi2201 studied in the present work reflects the
{\it intrinsic} behavior of the in-plane magnetic penetration
depth $\lambda_{ab}(T)$.

From the measured $\sigma_{sc}(T,H)$ we reconstructed $\lambda_{eff}^{-2}(T,H)$ by using the procedure described
in  Ref.~\onlinecite{Khasanov_08_InfLayer}. A correction between $\sigma_{sc}$
and $\lambda_{eff}^{-2}$ was considered according to:
\begin{equation}
\sigma_{sc}(b)[\mu{\rm s}^{-1}]=A(b)\lambda^{-2}_{eff}[{\rm nm}^{-2}],
 \label{eq:sigma-lambda}
\end{equation}
which accounts for decreasing of the field variance within the VL with increasing magnetic field.\cite{Brandt88,Brandt03} The correction factor $A(b)$ depends only on the reduced field $b=B/B_{c2}$ ($B_{c2}$ is the upper critical field). For  a superconductor with a Ginzburg-Landau parameter $\kappa=\lambda/\xi\geq5$  measured in fields ranging from $0.25/\kappa^{1.3}\lesssim b\leq1$, $A(b)$ can be obtained analytically as $A(b)=4.83\cdot10^4 (1 - b) [1 + 1.21(1 - \sqrt{b})^3]$~$\mu$s$^{-1}$nm$^{2}$ (see Ref.~\onlinecite{Brandt03}).

\begin{figure}[htb]
\includegraphics[width=0.8\linewidth]{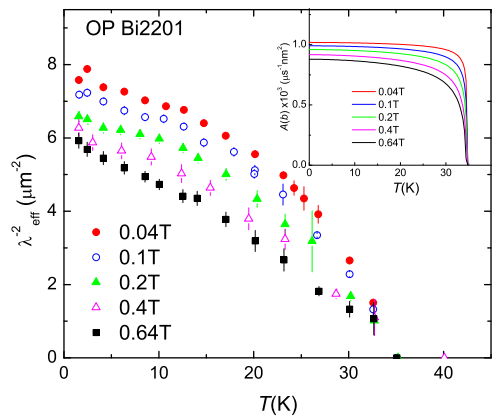}
\caption{(Color online) Temperature dependence  of
the effective magnetic penetration depth $\lambda_{eff}^{-2}$ reconstructed from
$\sigma_{sc}(T)$ measured at $\mu_0H=0.04$~T, 0.1~T, 0.2~T,
0.4~T, and 0.64~T (see Fig.~\ref{fig:sigma_vs_T}). The inset shows
the temperature dependence of the correction factor
$A(b)=\sigma_{sc}\cdot\lambda_{eff}^2$. }
 \label{fig:lambvda_eff}
\end{figure}

The reconstructed $\lambda_{eff}^{-2}(T,H={\rm const})$ curves are shown in Fig.~\ref{fig:lambvda_eff}. The corresponding $A(b)$ dependences are displayed in the inset.
The calculations were made for $B_{c2}(0)=50$~T.\cite{Wang03} The
temperature dependence of $B_{c2}$ was assumed to follow the
Werthamer-Helfand-Hohenberg (WHH) prediction.\cite{Werthamer66} Below $T\sim20$~K, $\lambda_{eff}^{-2}$ is linear in
$T$, as is expected for superconductor with nodes in the energy gap.
Fig.~\ref{fig:lambvda_eff} also implies that in the whole
temperature region (from $T\simeq1.6$~K up to $T_c$),
$\lambda_{eff}^{-2}(T,H)$ decreases with increasing field. This
contrasts the results obtained by using a similar procedure
for the ternary boride Li$_2$Pd$_3$B and electron-doped
Sr$_{0.9}$La$_{0.1}$CuO$_2$.\cite{Khasanov_08_InfLayer,Khasanov06_LiPdB} For these two
compounds the $\lambda_{eff}^{-2}(T,H)$ curves were found to collapse onto a single curve.
Since Li$_2$Pd$_3$B and Sr$_{0.9}$La$_{0.1}$CuO$_2$ are supposed to be fully gaped,
\cite{Khasanov_08_InfLayer,Khasanov06_LiPdB,Chen02,White08,Liu05,Hafliger08} we may conclude that the field dependence of $\lambda_{eff}^{-2}(T)$, shown in Fig.~\ref{fig:lambvda_eff}, is caused  by the presence of nodes in the superconducting energy gap of OP Bi2201.\cite{Kondo07}


%
\begin{figure}[htb]
\includegraphics[width=0.8\linewidth]{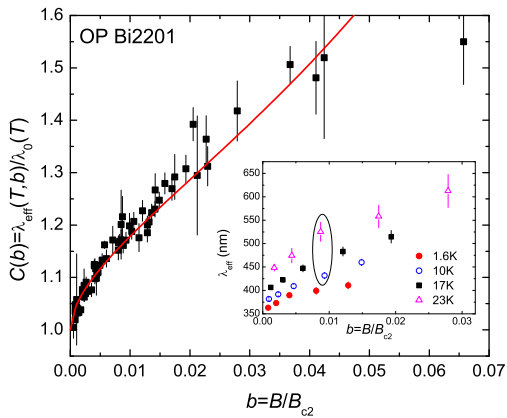}
\caption{(Color online) Dependence of the proportionality
factor $C(b)=\lambda_{eff}(T,b)/\lambda_0(T)$  on the reduced
field $b=B/B_{c2}$ of OP Bi2201. The solid line is the fit by means of
Eq.~(\ref{eq:nonlinear}) which takes into account the nonlinear
correction to $\lambda_0$. The inset shows the magnetic field dependence of
$\lambda_{eff}$ at $T=1.6$~K, 10~K, 17~K,
and 23~K. The reduced fields for two points selected in the oval
are almost the same.  }
 \label{fig:nonlinear-correction}
\end{figure}

As shown in Refs.~\onlinecite{Amin00} and \onlinecite{Amin99},
the magnetic field dependence of $\lambda_{eff}$ arises from the
nonlocal and the nonlinear response of a superconductor with nodes in
the energy gap to the applied magnetic field.
The nonlinear correction to $\lambda_{ab}$  appears due to the
magnetic field induced quasiparticle excitation over the gap nodes.\cite{Volovik93} According to Volovik,\cite{Volovik93} the
density of the delocalized states increases proportionally to
$\sqrt{b}$. The nonlocal correction to $\lambda_{ab}$ appears from the
response of electrons with momenta on the Fermi surface close to
the gap nodes. This is because the coherence length $\xi$, being
inversely proportional to the gap, becomes very large close to the
nodes and, formally, diverges at the nodal points. Thus there exist
areas on the Fermi surface where $\lambda/\xi\lesssim 1$, and
the response of a superconductor to an applied magnetic field
becomes highly nonlocal.\cite{Amin99}

In order to reconstruct the temperature dependence of the
superfluid density in zero magnetic field we used the general
assumption that the proportionality factor relating
$\lambda_{eff}$ to $\lambda_0$ is a function of the reduced
magnetic field $b$ only, so that:
\begin{equation}
\lambda_{eff}(b,T)=C(b)\lambda_0(T).
 \label{eq:lambda_eff}
\end{equation}
This statement is correct, at least, in case of nonlinear corrections which  scale with $\sqrt{b}$.\cite{Volovik93,Won01,Vekhter99}

\begin{figure}[htb]
\includegraphics[width=0.8\linewidth]{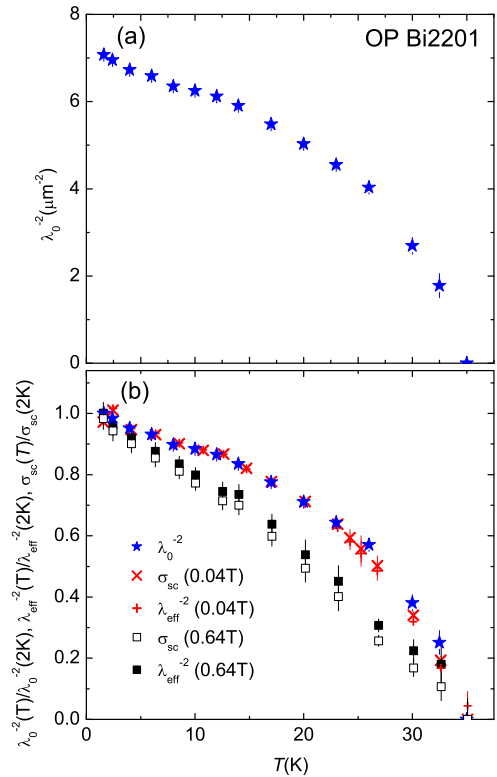}
\caption{(Color online) (a) Temperature dependence  of
$\lambda_{0}^{-2}$ of OP Bi2201 reconstructed from
$\lambda_{eff}^{-2}(T,H)$ measured at $\mu_0H=0.04$~T, 0.1~T,
0.2~T, 0.4~T, and 0.64~T (see Fig.~\ref{fig:lambvda_eff}).
(b) Temperature dependence of the normalized $\lambda_{0}^{-2}$,
$\lambda_{eff}^{-2}$(0.04~T), $\sigma_{sc}$(0.04~T),
$\lambda_{eff}^{-2}$(0.64~T), and $\sigma_{sc}$(0.64~T). }
 \label{fig:lambda_0}
\end{figure}

The fact that the field and the temperature dependences of
$\lambda_{eff}$ are described by separate terms [see  Eq.~(\ref{eq:lambda_eff})] allows to reconstruct
$\lambda_0(T)$. In order to demonstrate this, we refer to the inset in
Fig.~\ref{fig:nonlinear-correction} which shows the dependence of
$\lambda_{eff}$ on the magnetic field for some selected temperatures.
It is seen, {\it e.g.}, that the reduced field $b$ for
$\lambda_{eff}$ measured at $T=10$~K, $\mu_0H=0.4$~T (lower
point in the oval selection) is almost the same as the one for the point at
$T=23$~K, $\mu_0H=0.2$~T (upper point). This implies that the
coefficients $C(b)$ for these two points are nearly equal and that the
difference in the absolute values of $\lambda_{eff}$(10~K, 0.4~T)
and $\lambda_{eff}$(23~K, 0.2~T) is due to different values of
$\lambda_0$. The reconstruction procedure was performed in the
following way. First, from $\lambda_{eff}^{-2}(T,H)$ plotted in
Fig.~\ref{fig:lambvda_eff}, $\lambda_{eff}(b)$ was reconstructed for
various constant temperatures (see inset in
Fig.~\ref{fig:nonlinear-correction}). Second,
the resulting values of $\lambda_{eff}(T={\rm const},b)$  were scaled in order to have
them collapsing on a single curve (see
Fig.~\ref{fig:nonlinear-correction}). According to
Eq.~(\ref{eq:lambda_eff}) this curve corresponds to
$C(b)=\lambda_{eff}(T,b)/\lambda_0(T)$, while the scaling factor, in
turn, corresponds to $\lambda_0(T)$.
The solid line represents the result of the fit by means of the
relation:
\begin{equation}
\lambda_{eff}(b)/\lambda_0=C(b)=(1-K\sqrt{b})^{-1/2},
 \label{eq:nonlinear}
\end{equation}
which takes into account the nonlinear correction to $\lambda_0$ for
a  superconductor with $d-$wave energy gap.\cite{Vekhter99,Kadono04} Here the parameter $K$ depends on the
strength of the nonlinear effect. It is obvious that the
''nonlinear'' curve describes the experimental
$C(b)=\lambda_{eff}(b)/\lambda_0$ dependence reasonably well. In
particular, it reproduces the curvature at $b\lesssim0.01$ and the
linear increase of $C(b)$ for  $0.01\lesssim b \lesssim 0.05$. We
believe, however, that the whole $\lambda_{eff}(b)/\lambda_0$ curve
must be a combination of both nonlinear and nonlocal
correction effects, similar to the results of
Ref.~\onlinecite{Amin00}.

The temperature dependence of $\lambda_0^{-2}$ and the comparison of
$\lambda_0^{-2}(T)$ with $\sigma_{sc}(T)$ and
$\lambda_{eff}^{-2}(T)$ measured at $\mu_0H=0.04$~T and 0.64~T are
presented in Fig.~\ref{fig:lambda_0}. Both
$\sigma_{sc}(T)$ and $\lambda_{eff}^{-2}(T)$ measured at
$\mu_0H=0.04$~T almost coincide with each other as well as with
$\lambda_0^{-2}(T)$. This implies that the two sets of
corrections, namely, the first, accounting for decrease of the
second moment of the $\mu$SR line with increasing field
[Eq.~(\ref{eq:sigma-lambda}) and the inset in
Fig.~\ref{fig:lambvda_eff}] and the second, arising due to the
nonlocal and the nonlinear response of a superconductor with nodes
in the gap to the applied magnetic field [Eq.~(\ref{eq:lambda_eff})
and Fig.~\ref{fig:nonlinear-correction}], are not really important
at this relatively low field. Consequently, the second moment of
$\mu$SR line $\sigma_{sc}$ measured at $\mu_0H=0.04$~T is still a
good measure of the zero-field superfluid density $\rho_s\propto\lambda_0^{-2}$.
On the other hand, $\sigma_{sc}(T)$ and $\lambda_{eff}^{-2}(T)$ at
$\mu_0H=0.64$~T differ substantially from each other and from the
resulting $\lambda_0^{-2}(T)$. This implies that for high
fields both above mentioned corrections have to be taken
into account.


To conclude, muon-spin rotation measurements were performed on the optimally
doped cuprate superconductor
(BiPb)$_2$(SrLa)$_2$CuO$_{6+\delta}$. It was
demonstrated that in unconventional $d-$wave superconductors (like
OP Bi2201) $\mu$SR experiments taken in various
magnetic fields allow a reliable evaluation of the zero-field
superfluid density $\rho_s\propto\lambda_0^{-2}$. The deviation of the
effective penetration depth $\lambda_{eff}$ from $\lambda_0$ observed for higher fields was
explained by the nonlinear and the nonlocal response of the
superconductor with nodes in the energy gap to the applied
magnetic field. The dependence of $\lambda_{eff}/\lambda_0$ on the
reduced magnetic field $b=B/B_{c2}$ follows a $(1-K\sqrt{b})^{-1/2}$ behavior,
accounting for the nonlinear correction to $\lambda_0$ for a $d-$wave
superconductor.


This work was performed at the Swiss Muon Source (S$\mu$S), Paul
Scherrer Institute (PSI, Switzerland).
%
%
Work at the Ames Laboratory was supported by the Department of
Energy - Basic Energy Sciences under Contract No.
DE-AC02-07CH11358. The financial support of the Swiss National Foundation (SNF) is gratefully acknowledged.


\begin{thebibliography}{99}
%

\bibitem{Brandt88} E.H.~Brandt, Phys.~Rev.~B {\bf 37}, 2349
(1988).
%
\bibitem{Yaouanc03} A.~Yaouanc, P.~Dalmas~de~R\'{e}otier, and E.H.~Brandt, Phys.~Rev.~B {\bf 55}, 11107
(1997).
%
\bibitem{Brandt03} E.H.~Brandt, Phys.~Rev.~B {\bf 68}, 054506
(2003).
%
\bibitem{Sonier00} J.E.~Sonier, J.H.~Brewer, and R.F.~Kiefl, Rev.~Mod.~Phys. {\bf 72}, 769 (2000).
%
\bibitem{Sonier99} J.E.~Sonier, J.H.~Brewer, R.F.~Kiefl, G.D.~Morris, R.I.~Miller, D.A.~Bonn, J.~Chakhalian, R.H.~Heffner, W.N.~Hardy, and R.~Liang, Phys.~Rev.~Lett. {\bf 83}, 4156 (1999).
%
\bibitem{Kadono04} R.~Kadono, J.~Phys.:~Condens.~Matter {\bf 16}, S4421 (2004).
%
\bibitem{Niedermayer02} C.~Niedermayer, C.~Bernhard, T.~Holden, R.K.~Kremer, and K.~Ahn, Phys.~Rev.~B {\bf 65}, 094512 (2002).
%
\bibitem{Serventi04} S.~Serventi, G.~Allodi, R.~De~Renzi, G.~Guidi, L.~Romano, P.~Manfrinetti, A.~Palenzona, C.~Niedermayer, A.~Amato, and Ch.~Baines, Phys.~Rev.~Lett. {\bf 93}, 217003 (2004).
%
\bibitem{Khasanov07_La214} R.~Khasanov, A.~Shengelaya, A.~Maisuradze, F.~La~Mattina, A.~Bussmann-Holder, H.~Keller, and K.A.~M\"uller, Phys.~Rev.~Lett. {\bf 98}, 057007 (2007).
%
\bibitem{Khasanov07_Y123} R.~Khasanov, S.~Str\"assle, D.~Di~Castro, T.~Masui, S.~Miyasaka, S.~Tajima, A.~Bussmann-Holder, and H.~Keller, Phys.~Rev.~Lett. {\bf 99}, 237601 (2007).
%
\bibitem{Khasanov07_Y124} R.~Khasanov, A.~Shengelaya, A.~Bussmann-Holder, J.~Karpinski, H.~Keller, and  K.A.~M\"uller, J.~Supercond.~Nov.~Magn. {\bf 21}, 81 (2008).
%
\bibitem{Kondo04} T.~Kondo, T.~Takeuchi, T.~Yokoya, S.~Tsuda, S.~Shin, and U.~Mizutani,
J.~Electron~Spectrosc.~Relat.~Phenom. {\bf 137--140}, 663 (2004).
%
\bibitem{Kondo05} T.~Kondo, T.~Takeuchi, U.~Mizutani, T.~Yokoya, S.~Tsuda, and S.~Shin,
Phys.~Rev.~B {\bf 72}, 024533 (2005).
%
\bibitem{Khasanov08_Bi2201} R.~Khasanov, T.~Kondo, S.~Str\"assle, D.O.G.~Heron, A.~Kaminski,
H.~Keller, S.L.~Lee, and  T.~Takeuchi,  arXiv:0806.1907.
%
\bibitem{Lee93} S.L.~Lee, P.~Zimmermann, H.~Keller, M.~Warden, I.M.~Savi\'c, R.~Schauwecker, D.~Zech,  R.~Cubitt, E.M.~Forgan, P.H.~Kes, T.W.~Li, A.A.~Menovsky, and Z.~Tarnawski, Phys.~Rev.~Lett. {\bf 71}, 3862 (1993).
%
\bibitem{Aegerter98} C.M.~Aegerter, J.~Hofer, I.M.~Savi\'c, H.~Keller, S.L.~Lee, C.~Ager, S.H.~Lloyd, and E.M.~Forgan, Phys.~Rev.~B {\bf 57}, 1253 (1998).
%
\bibitem{Khasanov_08_InfLayer} R.~Khasanov, A.~Shengelaya, A.~Maisuradze, D.~Di~Castro, I.M.~Savi\'c, S.~Weyeneth, M.S.~Park, D.J.~Jang, S.-I.~Lee, and H.~Keller, Phys.~Rev.~B {\bf 77}, 184512 (2008).
%
\bibitem{Wang03} Y.~Wang, S.~Ono, Y.~Onose, G.~Gu, Y.~Ando, Y.~Tokura, S.~Uchida, and N.P.~Ong, Science {\bf 299}, 86 (2003).
%
\bibitem{Werthamer66}  N.R.~Werthamer, E.~Helfand, and P.C.~Hohenberg, Phys.~Rev. {\bf 147}, 295 (1966).
%
\bibitem{Khasanov06_LiPdB} R.~Khasanov, I.L.~Landau, C.~Baines, F.~La~Mattina, A.~Maisuradze, K.~Togano, and H. Keller, Phys.~Rev.~B {\bf 73}, 214528 (2006).
%
\bibitem{Chen02} C.-T.~Chen, P.~Seneor, N.-C.~Yeh, R.P.~Vasquez, L.D.~Bell, C.U.~Jung,  J.Y.~Kim, M.-S.~Park, H.-J.~Kim, and S.-I.~Lee, Phys.~Rev.~Lett. {\bf 88}, 227002 (2002).
%
\bibitem{White08} J.S.~White, E.M.~Forgan, M.~Laver,
P.S.~H\"afliger,  R.~Khasanov, R.~Cubitt, C.D.~Dewhurst,
M.S.~Park, D.-J.~Jang, and S.-I.~Lee, J.~Phys.:~Condens.~Matter.
{\bf 20}, 104237 (2008).
%
\bibitem{Liu05} Z.Y.~Liu, H.H.~Wen, L.~Shan, H.P~Yang, X.F.~Lu, H.~Gao,
M.-S.~Park, C.U.~Jung, and S.-I.~Lee,  Europhys.~Lett. {\bf 69}, 263 (2005).
%
\bibitem{Hafliger08}P.S.~H\"afliger, R.~Khasanov, R.~Lortz, A.~Petrovi\'c, K.~Togano, C.~Baines, B.~Graneli, and H.~Keller,  arXiv:0709.3777.
%
\bibitem{Kondo07} T.~Kondo, T.~Takeuchi, A.~Kaminski, S.~Tsuda, and S.~Shin, Phys.~Rev.~Lett. {\bf 98}, 267004 (2007).
%
\bibitem{Amin00} M.H.S.~Amin, M.~Franz, and I.~Affleck, Phys.~Rev.~Lett. {\bf 84},
5864 (2000).
%
\bibitem{Amin99} M.H.S.~Amin, Ph.D. thesis, University of
British Colombia (1999); cond-mat/0011455.
%
\bibitem{Volovik93} G.E.~Volovik, Sov.~Phys.~JETP~Lett. {\bf 58},
469 (1993).
%
\bibitem{Won01} H.~Won and K.~Maki, Europhys.~Lett. {\bf 54}, 248
(2001).
%
\bibitem{Vekhter99} I.~Vekhter, J.P.~Carbotte, and J.~Nicol,
Phys.~Rev.~B {\bf 59}, 1417 (1999).
%






\end{thebibliography}
\end{document}